\def\cp #1 #2 #3 {{\sl Chem.\ Phys.} {\bf #1}, #2 (#3)}
\def\jetp #1 #2 #3 {{\sl Sov.\ Phys.\ JETP} {\bf #1}, #2 (#3)}
\def\jfm #1 #2 #3 {{\sl J. Fluid\ Mech.} {\bf #1}, #2 (#3)}
\def\jpa #1 #2 #3 {{\sl J. Phys.\ A} {\bf #1}, #2 (#3)}
\def\jcp #1 #2 #3 {{\sl J.\ Chem.\ Phys.} {\bf #1}, #2 (#3)}
\def\jpc #1 #2 #3 {{\sl J.\ Phys.\ Chem.} {\bf #1}, #2 (#3)}
\def\jsp #1 #2 #3 {{\sl J.\ Stat.\ Phys.} {\bf #1}, #2 (#3)}
\def\jdep #1 #2 #3 {{\sl J.\ de Physique I} {\bf #1}, #2 (#3)}
\def\macro #1 #2 #3 {{\sl Macromolecules} {\bf #1}, #2 (#3)}
\def\pra #1 #2 #3 {{\sl Phys.\ Rev.\ A} {\bf #1}, #2 (#3)}
\def\prb #1 #2 #3 {{\sl Phys.\ Rev.\ B} {\bf #1}, #2 (#3)}
\def\pre #1 #2 #3 {{\sl Phys.\ Rev.\ E} {\bf #1}, #2 (#3)}
\def\prl #1 #2 #3 {{\sl Phys.\ Rev.\ Lett.} {\bf #1}, #2 (#3)}
\def\prsl #1 #2 #3 {{\sl Proc.\ Roy.\ Soc.\ London Ser. A} {\bf #1}, #2 (#3)}
\def\rmp #1 #2 #3 {{\sl Rev.\ Mod.\ Phys.} {\bf #1}, #2 (#3)}
\def\zpc #1 #2 #3 {{\sl Z. Phys.\ Chem.} {\bf #1}, #2 (#3)}
\def\zw #1 #2 #3 {{\sl Z. Wahrsch.\ verw.\ Gebiete} {\bf #1}, #2 (#3)}

\def\gtwid{\mathrel{\raise.3ex\hbox{$>$\kern-.75em\lower1ex\hbox{$\sim$}}}}
\def\ltwid{\mathrel{\raise.3ex\hbox{$<$\kern-.75em\lower1ex\hbox{$\sim$}}}}
\def\eg{{\it e.\ g.}}\def\ie{{\it i.\ e.}}\def\etc{{\it etc.}}
\def\vs{{\it vs.}}\def\vv{{\it vice versa}}\def\ea{{\it et al.}}
\def\apr{{\it a priori}}\def\apo{{\it a posteriori}}
\def\pd#1#2{{\partial #1\over\partial #2}}      
\def\p2d#1#2{{\partial^2 #1\over\partial #2^2}} 
\def\td#1#2{{d #1\over d #2}}      
\def\t2d#1#2{{d^2 #1\over d #2^2}} 
\def\ith{{$i^{\rm th}$}}
\def\jth{{$j^{\rm th}$}}
\def\kth{{$k^{\rm th}$}}
\def\nth{{$n^{\rm th}$}}
\def\dth{{$d^{\rm th}$}}
\def\eg{{\it e.\ g.}}\def\ie{{\it i.\ e.}}\def\etc{{\it etc.}}
\def\vs{{\it vs.}}\def\vv{{\it vice versa}}\def\ea{{\it et al.}}
\def\apr{{\it a priori}}\def\apo{{\it a posteriori}}

\newcount\refnum\refnum=0  
\def\refi{\smallskip\global\advance\refnum by 1\item{\the\refnum.}}
\newcount\rfignum\rfignum=0  
\def\rfigi{\medskip\global\advance\rfignum by 1\item{Figure \the\rfignum.}}

\newcount\eqnum \eqnum=0  
\def\eqnoi{\global\advance\eqnum by 1\eqno(\the\eqnum)}
\def\eqnai{\global\advance\eqnum by 1\eqno(\the\eqnum {\rm a})}
\def\eqnbi{\eqno(\the\eqnum {\rm b})}
\def\back#1{{\advance\eqnum by-#1 Eq.~(\the\eqnum)}}
\def\last{Eq.~(\the\eqnum)}                   

\def\pxt{P(x,t)}
\def\pyt{P(y,t)}
\def\pat{P(A,t)}
\def\pvt{P(V,t)}
\def\pxxt{P(x_1,x_2;t)}
\def\pix{P_0(x)}
\def\piy{P_0(y)}
\def\pixx{P_0(x_1,x_2)}
\def\pyyt{P(y_1,y_2;t)}
\def\msst{M(s_1,s_2;t)}
\def\mst{M(s,s;t)}
\def\mssi{M(s_1,s_2;0)}
\def\mss{M(s_1,s_2)}
\def\mss1{M(s_1+1,s_2+1)}
\def\mss1t{M(s_1+1,s_2+1;t)}
\def\mssst{M(s_1,\ldots,s_d;t)}
\def\dst{M(s,t)}
\def\ds1t{M(s+1,t)}
\def\dssst{M(s,\ldots,s;t)}
\def\ass{A(s_1,s_2)}
\def\ass1{A(s_1+1,s_2+1)}
\def\fiz{\Phi_1(z)}
\def\fz{\Phi_2(z)}
\def\Fz{\Phi_d(z)}
\def\fxt{\Phi_1(xt)}
\def\fAt{\Phi_2(At)}
\def\fVt{\Phi_d(Vt)}

\def\xniiav{\langle x_1^{n_1}\rangle}
\def\xniav{\langle x_1^n\rangle}
\def\xnfav{\langle x_2^n\rangle}
\def\xnffav{\langle x_2^{n_2}\rangle}
\def\Anav{\langle A^n\rangle}
\def\Vav{\langle V\rangle}
\def\Aav{\langle A\rangle}
\def\lav{\langle l\rangle}
\def\lnav{\langle l^n\rangle}
\def\ARnav{\langle (x_1/x_2)^n \rangle}
\def\xxnnav{\langle x_1^{n_1} x_2^{n_2} \rangle}
\def\xxnav{\langle (x_1x_2)^n\rangle}

\def\inx{{\int \limits_x^{\infty}dy}}
\def\ini{{\int \limits_0^{\infty}dy}}
\def\inii{{\int \limits_0^{\infty}\int \limits_0^{\infty} dx_1 dx_2}}
\def\inxx{{\int\limits_{x_1}^{\infty} \int\limits_{x_2}^{\infty} dy_1 dy_2}}

\def\a{\alpha}
\def\b{\beta}
\def\G{\Gamma}
\def\g{\gamma}
\def\d{\delta}
\def\z{\zeta}

\documentstyle[12pt]{article}
\renewcommand{\baselinestretch}{1.2}  

\begin{document}

\begin{titlepage}

\centerline{\bf Scaling and Multiscaling in Models of Fragmentation}
\bigskip
\centerline{\bf P.~L.~Krapivsky and E.~Ben-Naim\footnote{Permanent address:
The James Franck Institute, The University of Chicago,
5640 South Ellis Avenue, Chicago, IL 60637}}
\smallskip
\centerline{Center for Polymer Studies and Department of Physics}
\centerline{Boston University, Boston, MA 02215}
\vskip 1in

\centerline{\bf Abstract}
{\narrower\noindent We introduce a simple
geometric model which describes the kinetics of fragmentation of
$d$-dimensional objects. In one dimension our model coincides with the
random scission model and show a simple scaling behavior in the
long-time limit. For $d>1$, the volume of the fragments is characterized
by a single scale $1/t$, while other geometric properties such as
the length are characterized by an infinite number of length scales and
thus exhibit multiscaling.\bigskip}

{\noindent  P. A. C. S. Numbers: 02.50.-r, 05.40.+j, 82.20.-w, 82.20.Wt}
\bigskip
\end{titlepage}
\vfil\eject

\centerline{\bf I. Introduction}\bigskip

The phenomenon of fragmentation which occurs in numerous physical,
chemical, and geological processes, has attracted a considerable
recent interest. Fragmentation can be exemplified by polymer
degradation, grinding of minerals, atomic collisions cascades,
energy cascades in turbulence, multivalley structure of the phase space of
disorder systems, $\etc$ \cite{1,2,3,4,5,6,7}. In general, fragmentation is a
kinetic process with scattering, breaking, or splitting of
particular material into smaller fragments. With such wide-ranging
applications it is natural to try to abstract the essential features
of fragmentation and to model them as simple as possible.
One characteristic feature of these cascade processes
is that fragments continue splitting independently.
This allows one to describe the evolution by linear rate equations.
Another restriction which is used in almost all studies of
fragmentation is the implicit assumption that fragments may
be described by a single variable, say their mass or size.
The simplest model satisfying these restrictions is the so-called
random scission model \cite{5,8,9}. In this model,
the distribution of sizes is described by the integro-differential
equation,
\begin{equation}
\pd \pxt t = -x\pxt + 2\inx \pyt,
\label{1}
\end{equation}
where $\pxt$ is the concentration of fragments of mass (size) $x$,
$x$-mers, at time $t$. The loss term on the right-hand side represents
the decrease of $x$-mers due to binary breakups. The probability of
breaking at every point is assumed to be constant in the random
scission model and hence the overall rate at which an $x$-mer breaks
is equal to $x$. The gain term in Eq.~(\ref{1}) represents the increase of
$x$-mers due to breakups of longer fragments. The general solution
\cite{5,8,9} to Eq.~(\ref{1}) is
\begin{equation}
\pxt=e^{-xt}\left(\pix + \inx \piy[2t+t^2(y-x)]\right).
\label{2}
\end{equation}
In the long-time  limit,
this exact solution  approaches the scaling form
\begin{equation}
\pxt = Ct^2e^{-xt}, \quad C=\ini y\piy,
\label{3}
\end{equation}
if we keep $xt$ finite while taking a limit $t \to \infty$ and $x\to 0$.

The random scission model is a representative
example of ``one-dimensional'' fragmentation processes in which
the fragments are described by a single variable.
The kinetics of such fragmentation processes is now well
understood and numerous explicit and scaling solutions have
been found \cite{4,5,8,9,10,11,12,13,14}.

The geometry of fragments clearly influences the fragmentation
processes. However, it was ignored in so far studied models.  In this
paper, we introduce simple kinetic models describing the splitting of
two dimensional and more generally $d$-dimensional objects. We find that
multiscaling occurs for dimensions larger than one.
The rest of
this paper is organized as follows. In section II, we present the
two-dimensional model and analyze the behavior of the moments of the
size distribution of the fragments. Furthermore, we investigate
 the area distribution of the fragments an show that it
exhibits ordinary scaling. In Section III, we generalize the asymptotic
results to arbitrary dimensions and show that multiscaling occurs in
higher dimensions as well. In Section IV, we introduce a two-dimensional
isotropic fragmentation process. Numerical study of this process
suggests that it belongs to a different universality class.

\bigskip\centerline{\bf II. Fragmentation in two dimensions}\bigskip

In close analogy with the one-dimensional fragmentation process we
study the following process in two dimensions.
The fragmentation event takes place
at arbitrary internal point of the rectangular and gives birth
to four smaller rectangulars as illustrated in figure 1.
The distribution function $\pxxt$
describing rectangulars of size $x_1 \times x_2$,
is governed by the following kinetic equation,
\begin{equation}
\pd \pxxt t = -x_1x_2\pxt + 4\inxx \pyyt.
\label{4}
\end{equation}
Note that Eq.~(\ref{4}) implies the conservation of the total area,
\begin{equation}
\inii x_1x_2\pxxt = const.
\label{5}
\end{equation}

\begin{figure}[t]
\centerline{\begin{picture}(400,70)(30,705)
\thicklines
\put( 30,775){\line( 0,-1){ 70}}
\put( 30,775){\line( 1, 0){135}}
\put(165,775){\line( 0,-1){ 70}}
\put( 30,705){\line( 1, 0){135}}
\put(295,775){\line( 0,-1){ 70}}
\put(295,705){\line( 1, 0){135}}
\put(295,775){\line( 1, 0){135}}
\put(430,775){\line( 0,-1){ 70}}
\put(195,740){\vector( 1, 0){ 70}}
\put(395,775){\line( 0,-1){ 70}}
\put(295,725){\line( 1, 0){135}}
\end{picture}}
\caption[The Fragmentation Process]{The Fragmentation Process.}
\label{Fig-1}
\end{figure}
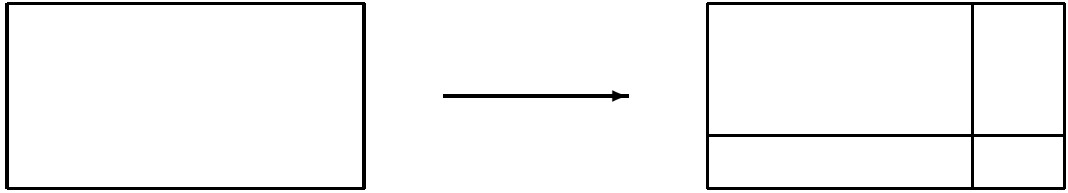

To analyze Eq.~(\ref{4}) we introduce the double Mellin transform of the
distribution function $\pxxt$,
\begin{equation}
\msst = \inii x_1^{s_1-1} x_2^{s_2-1}\pxxt.
\label{6}
\end{equation}
The functions $\msst$ at fixed $s_1$ and $s_2$
will be called the moments.
By combining Eqs.~(\ref{4}) and (\ref{6}) we arrive at the equation
\begin{equation}
\pd \msst t = \left({4 \over s_1s_2}-1\right)\mss1t.
\label{7}
\end{equation}

A surprising feature of Eq.~(\ref{7}) is that it implies the existence of
an infinite number of conservation laws. The moments $\msst$
with $s_1$ and $s_2$ satisfying the relation $s_1s_2=4$ are
independent of time. Thus in addition to the conservation of
the total area there is an infinite amount of
hidden conserved integrals. These integrals are in fact responsible for
the absence of scaling solutions to Eq.~(\ref{4}).
Indeed, with the scaling solution
$\pxxt=t^wQ(t^zx_1,t^zx_2)$, implies an infinite amount of
scaling relations, $w=z(s_1+s_2)$ at $s_1s_2=4$, which cannot
be satisfied by two scaling exponents, $w$ and $z$.

We will solve Eq.~(\ref{7}) by Charlesby's method \cite{8} (for more
recent applications of this method see, $\eg$, \cite{10,15}).  For the
random scission of the unit square, $P(x_1,x_2;0)=\d(x_1-1)\d(x_2-1)$,
or equivalently $\mssi=1$.  By iterating Eq.~(\ref{7}) one can compute
all derivatives of $\msst$ at $t=0$ and then find $\msst$ from the
Taylor's series, $\msst=M(0)+tM'(0)+t^2M''(0)/2!+t^3M'''(0)/3!+\ldots$.
This gives a solution in terms of a generalized hypergeometric function
\cite{16},
\begin{equation}
\msst= {}_2F_2(a_+,a_-;s_1,s_2;-t),
\label{8}
\end{equation}
with
\begin{equation}
a_{\pm} \equiv a_{\pm}(s_1,s_2)
={s_1+s_2\over 2}\pm\sqrt{\left({s_1-s_2 \over 2}\right)^2+4}.
\label{9}
\end{equation}

Computation of first few moments gives $M(1,1;t)\equiv N(t)=1+3t$
for the total number of fragments, $N(t)$; $M(2,2;t) \equiv 1$
for the total area; and $M(3,3;t)={1\over t}+{1\over 3t^2}
+e^{-t}\left({1\over 6} -{2\over 3t}+{1\over 3t^2}\right)$
for the next diagonal moment.  The first moment can be easily understood.
The rate of creation of fragments is equal to 3 since every
fragmentation event introduces 3 additional rectangulars,
and hence, the total number of fragments is $1+3t$.  These results
suggest the following power-law asymptotic behavior of the moments
$\msst$,
\begin{equation}
\msst \simeq A(s_1,s_2) t^{-\a(s_1,s_2)}.
\label{10}
\end{equation}
Substituting this asymptotic form into Eq.~(\ref{7}) yields the difference
equations for the exponent $\a(s_1,s_2)$, and for the prefactor $A(s_1,s_2)$,
\begin{eqnarray}
&\a(s_1,s_2)+1&=\a(s_1+1,s_2+1),\nonumber\\[-9pt]\\[-9pt]
&\a(s_1,s_2)A(s_1,s_2)&=\left(1-{4 \over s_1s_2}\right)A(s_1+1,s_2+1).\nonumber
\label{11}
\end{eqnarray}
With the boundary conditions, $\a(s_1,s_2)=0$ and $A(s_1,s_2)=1$ at
$s_1s_2=4$, Eq.~(11) are readily solved to give
\begin{eqnarray}
\label{12}
&\a(s_1,s_2)&=a_-(s_1,s_2)
={s_1+s_2\over 2}-\sqrt{\left({s_1-s_2 \over 2}\right)^2+4},
\nonumber\\[-9pt]\\[-9pt]
&A(s_1,s_2)&={\G(s_1)\G(s_2)\G(a_+-a_-)
\over \G(a_+-s_1)\G(a_+-s_2)\G(a_+)}.\nonumber
\end{eqnarray}
The  preceding formulas, Eqs.~(10) and (12), might be established
rigorously from the asymptotic behavior of the generalized
hypergeometric functions.

For ordinary scaling distributions the exponent $\a(s_1,s_2)$,
describing the asymptotic decay of the moments is linear in the
variable $s_1+s_2$. However, for two-dimensional fragmentation
 this exponent depends also on the variable $s_1-s_2$.
This manifests the non-trivial scaling properties of this process.
One can also compare the average value of $x_1^{n_1}x_2^{n_2}$,
defined by
\begin{equation}
\xxnnav={\inii x_1^{n_1} x_2^{n_2}\pxxt \over \inii \pxxt}
\equiv {M(n_1+1,n_2+1;t) \over M(1,1;t)},
\label{13}
\end{equation}
with the product $\xniiav \xnffav$. It turns out that the ratio of
these quantities {\it depends} asymptotically on time $t$, while for any
scaling distribution $\pxxt$ such a ratio would be a constant. In
particular,
\begin{equation}
{\xxnav \over \xniav \xnfav} \sim t^{-\left(\sqrt{n^2+16}-4\right)}.
\label{14}
\end{equation}
Only in the limit $n\to 0$ this ratio reaches a constant, while for every
positive $n$ the ratio decays in time.
By considering the case $n=1$
one sees that the average length, $\langle x_1\rangle\sim
t^{-(5-\sqrt{17})/2}\sim t^{-.438}$, decays slower than the square root
of the average area, $\sqrt{\langle x_1x_2\rangle} \sim t^{-1/2}$.
This again confirms that the fragment distribution function $\pxxt$
in the two-dimensional random scission model does not approach
a scaling form in the long-time limit. However, since all the moments
still show the power-law behavior we conclude that the model exhibits
a multiscaling asymptotic behavior.

The moments provide an almost complete analytical description
of the fragmentation process. However, the snapshot of the system
at the later stages remains intriguing (see Fig.~2).
This unexpectedly rich pattern arising in such a simple process
can be viewed as a consequence of the fact that the process
is not fully self-similar.  Instead, the pattern is formed of sets
of different scales which are spatially interwoven.
Fig.~2 also shows that a number of rectangulars have large
aspect ratio. This qualitative observation is in agreement
with the power-law behavior of the \nth\ moment of aspect ratio,
\begin{equation}
\ARnav \sim t^{\sqrt{n^2+4}-2},
\label{15}
\end{equation} which is valid for $|n|<1$.
Interestingly, this moment diverges when $|n|\ge1$. The aspect ratio
appears to be growing in time, in other words, perfect squares have great
tendency of breaking into long and tin rectangulars.

If we restrict ourselves to the {\it area} distribution function,
$\pat$,
\begin{equation}
\pat = \inii \d(x_1x_2-A)\pxxt,
\label{16}
\end{equation}
which provides a partial description of our system. We will show
that $\pat$ approaches the scaling form similar to those found
for other one-dimensional fragmentation systems \cite{9,10,12}.

\clearpage
\begin{figure}[t]
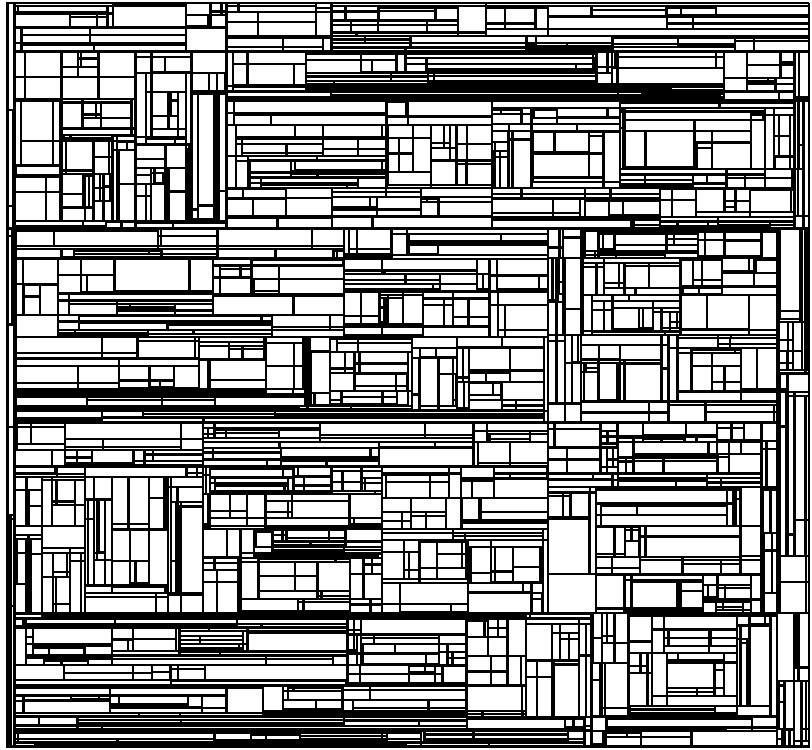

\setlength{\unitlength}{0.240900pt}
\ifx\plotpoint\undefined\newsavebox{\plotpoint}\fi
\sbox{\plotpoint}{\rule[-0.200pt]{0.400pt}{0.400pt}}%
\centerline{

}
\caption{Realization of the fragmentation process on a unit square at
  time t=1000.}
\label{Fig-2}
\end{figure}

Indeed,  the diagonal moments scale in time according to
\begin{equation}
M(s,s;t) \simeq {6\G(s) \over s(s+1)}t^{2-s},
\label{17}
\end{equation}
or in other words, the normalized \nth\ moments of the area
$\langle A^n\rangle^{1/n}$ are all proportional to $t^{-1}$.
Hence,  the area distribution function follows the scaling form
\begin{equation}
\pat \simeq t^2\fAt,
\label{18}
\end{equation}
where the scaling function $\fz$ satisfies
\begin{equation}
\int_0^{\infty}dz z^{s-1}\fz = {6\G(s) \over s(s+1)}.
\label{19}
\end{equation}
Performing the inverse Mellin transforms yields the
explicit expression for the scaling function \cite{17},
\begin{equation}
\fz = 6\int_0^1 d\z \left({1\over \z}-1\right)e^{-z/\z},
\label{20}
\end{equation}
with the limiting behavior
\begin{equation}
\fz \to  \cases {6 z^{-2}e^{-z},  &if $z \gg 1$,\cr
                 6\ln(1/z),       &if $z \ll 1$.\cr}
\label{21}
\end{equation}

Notice that the scaling solution of Eq.~(18) is characterized
by the same exponents as the scaling solution for the
one dimensional random scission model, $\pxt \sim t^2\fxt$.
However, the scaling functions are different, $\fiz=e^{-z}$
(see Eq.~(3)) is regular everywhere while $\fz$ diverges
logarithmically near the origin.

One can consider variations of this model for describing the kinetics
of fragmentation of multidimensional objects. For example,
one can change the governing rule of the fragmentation
events (see Fig.~1) and keep only two rectangulars,
say the rectangular in the bottom left corner and in the
upper right ones. This rule allows one to keep the total
length of fragments constant while the total area decays
to zero. Interestingly, this case has been partially
studied in connection with the problem of random sequential
adsorption of needles \cite{18}. This model can be treated by
applying our approach. One should just change in Eq.~(4)
the factor 4, corresponding to creation of four rectangulars,
by factor 2. Many results like Eqs.~(8), (10), and (12) remain
the same, with
\begin{equation}
a_{\pm}={s_1+s_2\over 2}\pm\sqrt{\left({s_1-s_2 \over 2}\right)^2+2},
\label{22}
\end{equation}
instead of Eq.~(9). All the qualitative conclusions also
do not change: the model exhibits a multiscaling asymptotic
behavior and, $\eg$,
\begin{equation}
{\xxnav \over \xniav \xnfav} \sim t^{-\left(\sqrt{n^2+8}-\sqrt{8}\right)}.
\label{23}
\end{equation}
The area distribution function again scales according to
\begin{equation}
\pat \simeq t^{\sqrt{2}}\fAt.
\label{24}
\end{equation}
Here the scaling function $\fz$ is given by
\begin{equation}
\fz = C\int_0^1{d\z\over \z}(1-\z)^{\sqrt{2}-1}e^{-z/\z},
\label{25}
\end{equation}
where $C={\G(2\sqrt{2})/\G^2(\sqrt{2})}=2.18482$.
In the limits of large and small area one has
\begin{equation}
\pat \to \cases {C A^{-2}e^{-At},            &if $At \gg 1$,\cr
                 D t^{\sqrt{2}}\ln(1/At),    &if $At \ll 1$,\cr}
\label{26}
\end{equation}
with $D={\G(2\sqrt{2})/\G^3(\sqrt{2})}=2.46432$.

\bigskip\centerline{\bf III. Generalization to Higher Dimensions}\bigskip

We turn now to the general $d$-dimensional random scission model.  The
asymptotic method presented for the two-dimensional case can be
generalized using a simple geometric construction. We are interested in
obtaining the moment $M({\bf s};t)$ with the notation ${\bf s}\equiv
(s_1,\ldots,s_n)$. In analogy with the two-dimensional case, we assume a
power-law behavior $M({\bf s};t)\sim t^{-\a(\bf s)}$. The exponents $\a$
satisfy the following difference equation
\begin{equation}
\a({\bf  s})+1=\a({\bf s}+{\bf 1}),\label{27}
\end{equation}
with the notation ${\bf 1}=(1,\ldots,1)$.
Meanwhile, the exponents
should also reflect the hidden conserved integrals, \ie, $\a({\bf s}^*)=0$
on the hypersurface ${\bf s}^*$, $\Pi_j s^*_j=2^d$.  The solution to
Eq.~(\ref{27}) with these boundary conditions is given by the formal expression
\begin{equation}
\a({\bf  s})=\a({\bf  s}^*+k{\bf 1})=k.  \label{28}
\end{equation}
This solution clearly satisfies the boundary condition as well as
Eq.~(28) .  Hence the problem is reduced to finding roots of
the algebraic equation $\Pi_j (s_j-k)=2^d$.
Since this equation is of degree $d$, a solution is  feasible only for
$d\le4$. An alternative way of viewing the solution is geometric. In
Eq.~(\ref{28}) ${\bf  s}^*+k{\bf 1}$ represents a line along the
$(1,\ldots,1)$ direction originating at ${\bf s^*}$ and ending at
${\bf s}={\bf  s}^*+k{\bf 1}$. The exponent $\a({\bf  s})$ equals the
projection of this line on an arbitrary axis, \eg\ ,
on the $s_1$ axis. Figure 3 illustrates this
construction.
\clearpage
\begin{figure}[t]
\setlength{\unitlength}{0.240900pt}
\ifx\plotpoint\undefined\newsavebox{\plotpoint}\fi
\sbox{\plotpoint}{\rule[-0.200pt]{0.400pt}{0.400pt}}%
\centerline{
\begin{picture}(1200,900)(0,0)
\font\gnuplot=cmr10 at 10pt
\gnuplot
\sbox{\plotpoint}{\rule[-0.500pt]{1.000pt}{1.000pt}}%
\put(220.0,113.0){\rule[-0.500pt]{220.664pt}{1.000pt}}
\put(220.0,113.0){\rule[-0.500pt]{1.000pt}{184.048pt}}
\put(220.0,113.0){\rule[-0.500pt]{220.664pt}{1.000pt}}
\put(1136.0,113.0){\rule[-0.500pt]{1.000pt}{184.048pt}}
\put(220.0,877.0){\rule[-0.500pt]{220.664pt}{1.000pt}}
\put(45,495){\makebox(0,0){$s_2$}}
\put(678,23){\makebox(0,0){$s_1$}}
\put(709,158){\makebox(0,0)[l]{$\,\alpha({\bf s})$}}
\put(611,279){\makebox(0,0)[l]{$\bf s^*$}}
\put(916,533){\makebox(0,0)[l]{$\bf s$}}
\put(220.0,113.0){\rule[-0.500pt]{1.000pt}{184.048pt}}
\put(611,126){\vector(1,0){305}}
\multiput(611.00,313.83)(0.599,0.500){500}{\rule{1.451pt}{0.120pt}}
\multiput(611.00,309.92)(301.989,254.000){2}{\rule{0.725pt}{1.000pt}}
\put(916,566){\vector(4,3){0}}
\put(322,877){\usebox{\plotpoint}}
\multiput(323.83,846.44)(0.485,-3.716){10}{\rule{0.117pt}{7.361pt}}
\multiput(319.92,861.72)(9.000,-48.722){2}{\rule{1.000pt}{3.681pt}}
\multiput(332.83,787.06)(0.485,-3.117){10}{\rule{0.117pt}{6.250pt}}
\multiput(328.92,800.03)(9.000,-41.028){2}{\rule{1.000pt}{3.125pt}}
\multiput(341.83,738.87)(0.487,-2.352){12}{\rule{0.117pt}{4.850pt}}
\multiput(337.92,748.93)(10.000,-35.934){2}{\rule{1.000pt}{2.425pt}}
\multiput(351.83,693.51)(0.485,-2.279){10}{\rule{0.117pt}{4.694pt}}
\multiput(347.92,703.26)(9.000,-30.256){2}{\rule{1.000pt}{2.347pt}}
\multiput(360.83,655.82)(0.485,-1.980){10}{\rule{0.117pt}{4.139pt}}
\multiput(356.92,664.41)(9.000,-26.410){2}{\rule{1.000pt}{2.069pt}}
\multiput(369.83,622.66)(0.485,-1.740){10}{\rule{0.117pt}{3.694pt}}
\multiput(365.92,630.33)(9.000,-23.332){2}{\rule{1.000pt}{1.847pt}}
\multiput(378.83,594.75)(0.487,-1.341){12}{\rule{0.117pt}{2.950pt}}
\multiput(374.92,600.88)(10.000,-20.877){2}{\rule{1.000pt}{1.475pt}}
\multiput(388.83,567.43)(0.485,-1.381){10}{\rule{0.117pt}{3.028pt}}
\multiput(384.92,573.72)(9.000,-18.716){2}{\rule{1.000pt}{1.514pt}}
\multiput(397.83,543.82)(0.485,-1.202){10}{\rule{0.117pt}{2.694pt}}
\multiput(393.92,549.41)(9.000,-16.408){2}{\rule{1.000pt}{1.347pt}}
\multiput(406.83,522.74)(0.485,-1.082){10}{\rule{0.117pt}{2.472pt}}
\multiput(402.92,527.87)(9.000,-14.869){2}{\rule{1.000pt}{1.236pt}}
\multiput(415.83,504.49)(0.487,-0.863){12}{\rule{0.117pt}{2.050pt}}
\multiput(411.92,508.75)(10.000,-13.745){2}{\rule{1.000pt}{1.025pt}}
\multiput(425.83,486.12)(0.485,-0.902){10}{\rule{0.117pt}{2.139pt}}
\multiput(421.92,490.56)(9.000,-12.561){2}{\rule{1.000pt}{1.069pt}}
\multiput(434.83,470.04)(0.485,-0.783){10}{\rule{0.117pt}{1.917pt}}
\multiput(430.92,474.02)(9.000,-11.022){2}{\rule{1.000pt}{0.958pt}}
\multiput(443.83,455.50)(0.485,-0.723){10}{\rule{0.117pt}{1.806pt}}
\multiput(439.92,459.25)(9.000,-10.252){2}{\rule{1.000pt}{0.903pt}}
\multiput(452.83,442.57)(0.487,-0.597){12}{\rule{0.117pt}{1.550pt}}
\multiput(448.92,445.78)(10.000,-9.783){2}{\rule{1.000pt}{0.775pt}}
\multiput(462.83,429.43)(0.485,-0.603){10}{\rule{0.117pt}{1.583pt}}
\multiput(458.92,432.71)(9.000,-8.714){2}{\rule{1.000pt}{0.792pt}}
\multiput(471.83,417.89)(0.485,-0.543){10}{\rule{0.117pt}{1.472pt}}
\multiput(467.92,420.94)(9.000,-7.944){2}{\rule{1.000pt}{0.736pt}}
\multiput(480.83,407.35)(0.485,-0.483){10}{\rule{0.117pt}{1.361pt}}
\multiput(476.92,410.17)(9.000,-7.175){2}{\rule{1.000pt}{0.681pt}}
\multiput(488.00,400.68)(0.437,-0.487){12}{\rule{1.250pt}{0.117pt}}
\multiput(488.00,400.92)(7.406,-10.000){2}{\rule{0.625pt}{1.000pt}}
\multiput(498.00,390.68)(0.423,-0.485){10}{\rule{1.250pt}{0.117pt}}
\multiput(498.00,390.92)(6.406,-9.000){2}{\rule{0.625pt}{1.000pt}}
\multiput(507.00,381.68)(0.470,-0.481){8}{\rule{1.375pt}{0.116pt}}
\multiput(507.00,381.92)(6.146,-8.000){2}{\rule{0.688pt}{1.000pt}}
\multiput(516.00,373.68)(0.470,-0.481){8}{\rule{1.375pt}{0.116pt}}
\multiput(516.00,373.92)(6.146,-8.000){2}{\rule{0.688pt}{1.000pt}}
\multiput(525.00,365.68)(0.539,-0.481){8}{\rule{1.500pt}{0.116pt}}
\multiput(525.00,365.92)(6.887,-8.000){2}{\rule{0.750pt}{1.000pt}}
\multiput(535.00,357.69)(0.525,-0.475){6}{\rule{1.536pt}{0.114pt}}
\multiput(535.00,357.92)(5.813,-7.000){2}{\rule{0.768pt}{1.000pt}}
\multiput(544.00,350.69)(0.525,-0.475){6}{\rule{1.536pt}{0.114pt}}
\multiput(544.00,350.92)(5.813,-7.000){2}{\rule{0.768pt}{1.000pt}}
\multiput(553.00,343.69)(0.579,-0.462){4}{\rule{1.750pt}{0.111pt}}
\multiput(553.00,343.92)(5.368,-6.000){2}{\rule{0.875pt}{1.000pt}}
\multiput(562.00,337.69)(0.681,-0.462){4}{\rule{1.917pt}{0.111pt}}
\multiput(562.00,337.92)(6.022,-6.000){2}{\rule{0.958pt}{1.000pt}}
\multiput(572.00,331.69)(0.579,-0.462){4}{\rule{1.750pt}{0.111pt}}
\multiput(572.00,331.92)(5.368,-6.000){2}{\rule{0.875pt}{1.000pt}}
\multiput(581.00,325.71)(0.490,-0.424){2}{\rule{2.050pt}{0.102pt}}
\multiput(581.00,325.92)(4.745,-5.000){2}{\rule{1.025pt}{1.000pt}}
\multiput(590.00,320.71)(0.490,-0.424){2}{\rule{2.050pt}{0.102pt}}
\multiput(590.00,320.92)(4.745,-5.000){2}{\rule{1.025pt}{1.000pt}}
\multiput(599.00,315.71)(0.660,-0.424){2}{\rule{2.250pt}{0.102pt}}
\multiput(599.00,315.92)(5.330,-5.000){2}{\rule{1.125pt}{1.000pt}}
\multiput(609.00,310.71)(0.490,-0.424){2}{\rule{2.050pt}{0.102pt}}
\multiput(609.00,310.92)(4.745,-5.000){2}{\rule{1.025pt}{1.000pt}}
\put(618,303.92){\rule{2.168pt}{1.000pt}}
\multiput(618.00,305.92)(4.500,-4.000){2}{\rule{1.084pt}{1.000pt}}
\put(627,299.92){\rule{2.168pt}{1.000pt}}
\multiput(627.00,301.92)(4.500,-4.000){2}{\rule{1.084pt}{1.000pt}}
\put(636,295.92){\rule{2.409pt}{1.000pt}}
\multiput(636.00,297.92)(5.000,-4.000){2}{\rule{1.204pt}{1.000pt}}
\put(646,291.92){\rule{2.168pt}{1.000pt}}
\multiput(646.00,293.92)(4.500,-4.000){2}{\rule{1.084pt}{1.000pt}}
\put(655,287.92){\rule{2.168pt}{1.000pt}}
\multiput(655.00,289.92)(4.500,-4.000){2}{\rule{1.084pt}{1.000pt}}
\put(664,284.42){\rule{2.168pt}{1.000pt}}
\multiput(664.00,285.92)(4.500,-3.000){2}{\rule{1.084pt}{1.000pt}}
\put(673,280.92){\rule{2.409pt}{1.000pt}}
\multiput(673.00,282.92)(5.000,-4.000){2}{\rule{1.204pt}{1.000pt}}
\put(683,277.42){\rule{2.168pt}{1.000pt}}
\multiput(683.00,278.92)(4.500,-3.000){2}{\rule{1.084pt}{1.000pt}}
\put(692,274.42){\rule{2.168pt}{1.000pt}}
\multiput(692.00,275.92)(4.500,-3.000){2}{\rule{1.084pt}{1.000pt}}
\put(701,271.42){\rule{2.168pt}{1.000pt}}
\multiput(701.00,272.92)(4.500,-3.000){2}{\rule{1.084pt}{1.000pt}}
\put(710,268.42){\rule{2.409pt}{1.000pt}}
\multiput(710.00,269.92)(5.000,-3.000){2}{\rule{1.204pt}{1.000pt}}
\put(720,265.42){\rule{2.168pt}{1.000pt}}
\multiput(720.00,266.92)(4.500,-3.000){2}{\rule{1.084pt}{1.000pt}}
\put(729,262.42){\rule{2.168pt}{1.000pt}}
\multiput(729.00,263.92)(4.500,-3.000){2}{\rule{1.084pt}{1.000pt}}
\put(738,259.42){\rule{2.168pt}{1.000pt}}
\multiput(738.00,260.92)(4.500,-3.000){2}{\rule{1.084pt}{1.000pt}}
\put(747,256.92){\rule{2.409pt}{1.000pt}}
\multiput(747.00,257.92)(5.000,-2.000){2}{\rule{1.204pt}{1.000pt}}
\put(757,254.42){\rule{2.168pt}{1.000pt}}
\multiput(757.00,255.92)(4.500,-3.000){2}{\rule{1.084pt}{1.000pt}}
\put(766,251.92){\rule{2.168pt}{1.000pt}}
\multiput(766.00,252.92)(4.500,-2.000){2}{\rule{1.084pt}{1.000pt}}
\put(775,249.92){\rule{2.168pt}{1.000pt}}
\multiput(775.00,250.92)(4.500,-2.000){2}{\rule{1.084pt}{1.000pt}}
\put(784,247.92){\rule{2.409pt}{1.000pt}}
\multiput(784.00,248.92)(5.000,-2.000){2}{\rule{1.204pt}{1.000pt}}
\put(794,245.42){\rule{2.168pt}{1.000pt}}
\multiput(794.00,246.92)(4.500,-3.000){2}{\rule{1.084pt}{1.000pt}}
\put(803,242.92){\rule{2.168pt}{1.000pt}}
\multiput(803.00,243.92)(4.500,-2.000){2}{\rule{1.084pt}{1.000pt}}
\put(812,240.92){\rule{2.168pt}{1.000pt}}
\multiput(812.00,241.92)(4.500,-2.000){2}{\rule{1.084pt}{1.000pt}}
\put(821,238.92){\rule{2.409pt}{1.000pt}}
\multiput(821.00,239.92)(5.000,-2.000){2}{\rule{1.204pt}{1.000pt}}
\put(831,236.92){\rule{2.168pt}{1.000pt}}
\multiput(831.00,237.92)(4.500,-2.000){2}{\rule{1.084pt}{1.000pt}}
\put(840,235.42){\rule{2.168pt}{1.000pt}}
\multiput(840.00,235.92)(4.500,-1.000){2}{\rule{1.084pt}{1.000pt}}
\put(849,233.92){\rule{2.168pt}{1.000pt}}
\multiput(849.00,234.92)(4.500,-2.000){2}{\rule{1.084pt}{1.000pt}}
\put(858,231.92){\rule{2.409pt}{1.000pt}}
\multiput(858.00,232.92)(5.000,-2.000){2}{\rule{1.204pt}{1.000pt}}
\put(868,229.92){\rule{2.168pt}{1.000pt}}
\multiput(868.00,230.92)(4.500,-2.000){2}{\rule{1.084pt}{1.000pt}}
\put(877,228.42){\rule{2.168pt}{1.000pt}}
\multiput(877.00,228.92)(4.500,-1.000){2}{\rule{1.084pt}{1.000pt}}
\put(886,226.92){\rule{2.168pt}{1.000pt}}
\multiput(886.00,227.92)(4.500,-2.000){2}{\rule{1.084pt}{1.000pt}}
\put(895,225.42){\rule{2.409pt}{1.000pt}}
\multiput(895.00,225.92)(5.000,-1.000){2}{\rule{1.204pt}{1.000pt}}
\put(905,223.92){\rule{2.168pt}{1.000pt}}
\multiput(905.00,224.92)(4.500,-2.000){2}{\rule{1.084pt}{1.000pt}}
\put(914,222.42){\rule{2.168pt}{1.000pt}}
\multiput(914.00,222.92)(4.500,-1.000){2}{\rule{1.084pt}{1.000pt}}
\put(923,220.92){\rule{2.168pt}{1.000pt}}
\multiput(923.00,221.92)(4.500,-2.000){2}{\rule{1.084pt}{1.000pt}}
\put(932,219.42){\rule{2.409pt}{1.000pt}}
\multiput(932.00,219.92)(5.000,-1.000){2}{\rule{1.204pt}{1.000pt}}
\put(942,217.92){\rule{2.168pt}{1.000pt}}
\multiput(942.00,218.92)(4.500,-2.000){2}{\rule{1.084pt}{1.000pt}}
\put(951,216.42){\rule{2.168pt}{1.000pt}}
\multiput(951.00,216.92)(4.500,-1.000){2}{\rule{1.084pt}{1.000pt}}
\put(960,215.42){\rule{2.168pt}{1.000pt}}
\multiput(960.00,215.92)(4.500,-1.000){2}{\rule{1.084pt}{1.000pt}}
\put(969,213.92){\rule{2.409pt}{1.000pt}}
\multiput(969.00,214.92)(5.000,-2.000){2}{\rule{1.204pt}{1.000pt}}
\put(979,212.42){\rule{2.168pt}{1.000pt}}
\multiput(979.00,212.92)(4.500,-1.000){2}{\rule{1.084pt}{1.000pt}}
\put(988,211.42){\rule{2.168pt}{1.000pt}}
\multiput(988.00,211.92)(4.500,-1.000){2}{\rule{1.084pt}{1.000pt}}
\put(997,210.42){\rule{2.168pt}{1.000pt}}
\multiput(997.00,210.92)(4.500,-1.000){2}{\rule{1.084pt}{1.000pt}}
\put(1006,209.42){\rule{2.409pt}{1.000pt}}
\multiput(1006.00,209.92)(5.000,-1.000){2}{\rule{1.204pt}{1.000pt}}
\put(1016,208.42){\rule{2.168pt}{1.000pt}}
\multiput(1016.00,208.92)(4.500,-1.000){2}{\rule{1.084pt}{1.000pt}}
\put(1025,207.42){\rule{2.168pt}{1.000pt}}
\multiput(1025.00,207.92)(4.500,-1.000){2}{\rule{1.084pt}{1.000pt}}
\put(1034,205.92){\rule{2.168pt}{1.000pt}}
\multiput(1034.00,206.92)(4.500,-2.000){2}{\rule{1.084pt}{1.000pt}}
\put(1043,204.42){\rule{2.409pt}{1.000pt}}
\multiput(1043.00,204.92)(5.000,-1.000){2}{\rule{1.204pt}{1.000pt}}
\put(1053,203.42){\rule{2.168pt}{1.000pt}}
\multiput(1053.00,203.92)(4.500,-1.000){2}{\rule{1.084pt}{1.000pt}}
\put(1062,202.42){\rule{2.168pt}{1.000pt}}
\multiput(1062.00,202.92)(4.500,-1.000){2}{\rule{1.084pt}{1.000pt}}
\put(1071,201.42){\rule{2.168pt}{1.000pt}}
\multiput(1071.00,201.92)(4.500,-1.000){2}{\rule{1.084pt}{1.000pt}}
\put(1080,200.42){\rule{2.409pt}{1.000pt}}
\multiput(1080.00,200.92)(5.000,-1.000){2}{\rule{1.204pt}{1.000pt}}
\put(1090,199.42){\rule{2.168pt}{1.000pt}}
\multiput(1090.00,199.92)(4.500,-1.000){2}{\rule{1.084pt}{1.000pt}}
\put(1108,198.42){\rule{2.168pt}{1.000pt}}
\multiput(1108.00,198.92)(4.500,-1.000){2}{\rule{1.084pt}{1.000pt}}
\put(1117,197.42){\rule{2.409pt}{1.000pt}}
\multiput(1117.00,197.92)(5.000,-1.000){2}{\rule{1.204pt}{1.000pt}}
\put(1127,196.42){\rule{2.168pt}{1.000pt}}
\multiput(1127.00,196.92)(4.500,-1.000){2}{\rule{1.084pt}{1.000pt}}
\put(1099.0,201.0){\rule[-0.500pt]{2.168pt}{1.000pt}}
\put(611,312){\raisebox{-.8pt}{\makebox(0,0){$\Diamond$}}}
\put(916,566){\raisebox{-.8pt}{\makebox(0,0){$\Diamond$}}}
\end{picture}
}
\caption[Geometric Solution]{The Geometric Solution. The hypersurface
${\bf s}^*$ satisfies $\Pi_j s_j=2^d$.}
\label{Fig-3}
\end{figure}
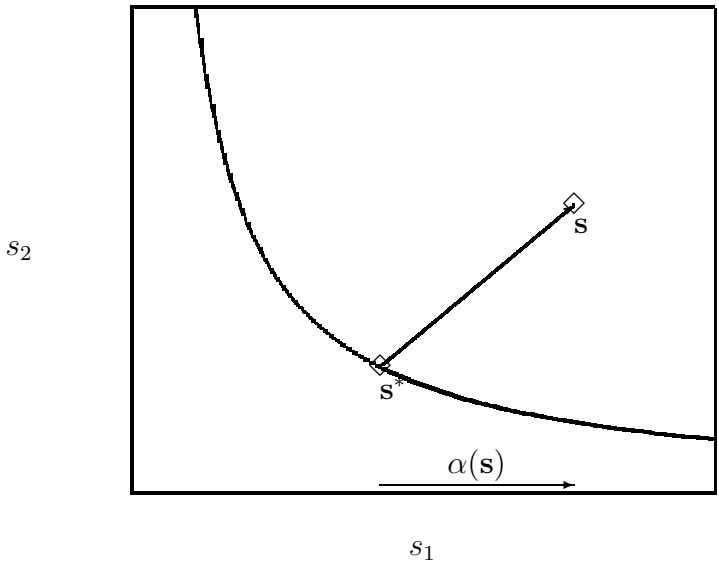

The main features found for the two-dimensional case such as
multiscaling occur for higher dimensions as well. As a manifestation of
the existence of multiple length scales in the system let us consider
the ratio of the average volume $\Vav$ to the \dth\ power of the average
length, $\lav$. We define the exponent $\beta_d$ by
\begin{equation}
{\Vav\over\lav}\sim t^{-\beta_d},\label{29}
\end{equation}
or equivalently, $\beta_d=1-d\left(\alpha(2,1\ldots,1)+1\right)$. Using
the construction of Eq.~(\ref{27}) , we find $\beta_d=0,0.1231,0.1486$
for $d=1,2,3$ respectively, while in the limit $d\to\infty$ this
exponent saturates at $1-2\log(3/2)\cong0.1891$.  Note that $\beta_d$
measures the deviation between the asymptotic behavior of the length
and the volume. As the dimension is increased, this discrepancy becomes
more pronounced,  and hence, multiscaling is stronger in higher
dimensions. Another consequence of the same phenomenon is the
nonuniversal behavior of the various moments of the length
distribution. We find that the \nth\ moment decays asymptotically
according to
\begin{equation}
\langle l^n\rangle\sim t^{-2\log(1+n/2)/d},\label{30}
\end{equation}
indicating the presence of an infinite number of length scales.

One can also show that different directions behave independently to a
certain degree in the limit of infinite dimensions. Specifically, one
can show that
\begin{equation}
\left\langle \Pi_j x_j^{n_j}\right\rangle=
\Pi_j \left\langle  x_j^{n_j}\right\rangle, \label{31}
\end{equation}
if $n_j=0$ except for a finite number. The average over a finite number
of variables decouples into a product over single variable averages,
while the average over an infinite number does decouple.

For completeness, we present the general dimension results for the
diagonal moments, $\dssst$, which will be shortly denoted by $\dst$.
The governing equation for these moments reads
\begin{equation}
\pd \dst t = \left[(2/s)^d-1\right]\ds1t.
\label{32}
\end{equation}
We substitute the power-law asymptotic behavior,
$\dst \simeq A(s) t^{-\a(s)}$, into Eq.~(32) and  take into account
the boundary conditions $\a(s=2)=0$ and $A(s=2)=1$.
By solving the resulting difference equations we find
\begin{equation}
\a(s)=s-2, \quad
A(s)=\G^d(s)\prod_{j=1}^{d-1}{\G(2-2\cdot\z^j) \over\G(s-2\cdot\z^j)},
\label{33}
\end{equation}
with $\z=\exp(2\pi i/d)$.

In the long-time limit the volume
distribution function, $\pvt$, approaches the scaling form
\begin{equation}
\pvt \simeq t^2\fVt,
\label{34}
\end{equation}
with $\Fz$ being the inverse Mellon transform of $A(s)$.
After a lengthy calculation one can find the asymptotic behavior
of the volume distribution function:
\begin{equation}
\pvt \to \cases {C_d V^{-2}e^{-Vt},            &if $Vt \gg 1$,\cr
                 D_d t^2\ln^{d-1}(1/Vt),       &if $Vt \ll 1$,\cr}
\label{35}
\end{equation}
where $C_d=\prod_{j=1}^{d-1}\G(2-2\cdot\z^j)$ and
$D_d=2^{d-1}(2^d-1)/\G(d)$. Thus for all $d>1$ the volume distribution
function diverges logarithmically in the small-volume limit.

To summarize, in the $d$-dimensional random scission model,
the volume is characterized by only one scale, $V \sim t^{-1}$.
However, other geometrical characteristics such as the average
length and the surface area decay nonuniversaly in time because
of the existence of an infinite amount of length scales,
namely multiscaling.

One can also consider a varying fragmentation rate and study
the case when the overall rate depends on the volume as a power-law,
$\ie$, as $V^{\lambda}$ (the case $\lambda=1$ corresponds
to the random scission model).
When $\lambda$ is positive, this generalization also results in
multiscaling of the fragments distribution.
The total number of fragments $N(t)\equiv M({\bf 1};t)$ grows
algebraically in time, $N(t)\sim t^{1/\lambda}$.
Hence, the case $\lambda=0$ is a critical one and the
number of fragments grows exponentially in time.
Finally, for $\lambda<0$ the shattering transition takes place:
the total volume decreases monotonically and the total number
of fragments reaches infinity within an infinitesimally small
time interval. Moreover, a finite fraction of the volume breaks into
zero-volume rectangulars. This phenomenon is well known in the
context of one-dimensional fragmentation \cite{10,11} and has been
examined in the context of two-dimensional fragmentation
with length conservation in a very recent study \cite{19}.

\bigskip\centerline{\bf IV. Isotropic Fragmentation}\bigskip

\begin{figure}[t]
\centerline{
\setlength{\unitlength}{0.240900pt}
\ifx\plotpoint\undefined\newsavebox{\plotpoint}\fi
\sbox{\plotpoint}{\rule[-0.500pt]{1.000pt}{1.000pt}}%
\begin{picture}(1500,900)(0,0)
\font\gnuplot=cmr10 at 10pt
\gnuplot
\sbox{\plotpoint}{\rule[-0.500pt]{1.000pt}{1.000pt}}%
\put(220.0,113.0){\rule[-0.500pt]{4.818pt}{1.000pt}}
\put(198,113){\makebox(0,0)[r]{$0.01$}}
\put(1416.0,113.0){\rule[-0.500pt]{4.818pt}{1.000pt}}
\put(220.0,228.0){\rule[-0.500pt]{2.409pt}{1.000pt}}
\put(1426.0,228.0){\rule[-0.500pt]{2.409pt}{1.000pt}}
\put(220.0,295.0){\rule[-0.500pt]{2.409pt}{1.000pt}}
\put(1426.0,295.0){\rule[-0.500pt]{2.409pt}{1.000pt}}
\put(220.0,343.0){\rule[-0.500pt]{2.409pt}{1.000pt}}
\put(1426.0,343.0){\rule[-0.500pt]{2.409pt}{1.000pt}}
\put(220.0,380.0){\rule[-0.500pt]{2.409pt}{1.000pt}}
\put(1426.0,380.0){\rule[-0.500pt]{2.409pt}{1.000pt}}
\put(220.0,410.0){\rule[-0.500pt]{2.409pt}{1.000pt}}
\put(1426.0,410.0){\rule[-0.500pt]{2.409pt}{1.000pt}}
\put(220.0,436.0){\rule[-0.500pt]{2.409pt}{1.000pt}}
\put(1426.0,436.0){\rule[-0.500pt]{2.409pt}{1.000pt}}
\put(220.0,458.0){\rule[-0.500pt]{2.409pt}{1.000pt}}
\put(1426.0,458.0){\rule[-0.500pt]{2.409pt}{1.000pt}}
\put(220.0,478.0){\rule[-0.500pt]{2.409pt}{1.000pt}}
\put(1426.0,478.0){\rule[-0.500pt]{2.409pt}{1.000pt}}
\put(220.0,495.0){\rule[-0.500pt]{4.818pt}{1.000pt}}
\put(198,495){\makebox(0,0)[r]{$0.1$}}
\put(1416.0,495.0){\rule[-0.500pt]{4.818pt}{1.000pt}}
\put(220.0,610.0){\rule[-0.500pt]{2.409pt}{1.000pt}}
\put(1426.0,610.0){\rule[-0.500pt]{2.409pt}{1.000pt}}
\put(220.0,677.0){\rule[-0.500pt]{2.409pt}{1.000pt}}
\put(1426.0,677.0){\rule[-0.500pt]{2.409pt}{1.000pt}}
\put(220.0,725.0){\rule[-0.500pt]{2.409pt}{1.000pt}}
\put(1426.0,725.0){\rule[-0.500pt]{2.409pt}{1.000pt}}
\put(220.0,762.0){\rule[-0.500pt]{2.409pt}{1.000pt}}
\put(1426.0,762.0){\rule[-0.500pt]{2.409pt}{1.000pt}}
\put(220.0,792.0){\rule[-0.500pt]{2.409pt}{1.000pt}}
\put(1426.0,792.0){\rule[-0.500pt]{2.409pt}{1.000pt}}
\put(220.0,818.0){\rule[-0.500pt]{2.409pt}{1.000pt}}
\put(1426.0,818.0){\rule[-0.500pt]{2.409pt}{1.000pt}}
\put(220.0,840.0){\rule[-0.500pt]{2.409pt}{1.000pt}}
\put(1426.0,840.0){\rule[-0.500pt]{2.409pt}{1.000pt}}
\put(220.0,860.0){\rule[-0.500pt]{2.409pt}{1.000pt}}
\put(1426.0,860.0){\rule[-0.500pt]{2.409pt}{1.000pt}}
\put(220.0,877.0){\rule[-0.500pt]{4.818pt}{1.000pt}}
\put(198,877){\makebox(0,0)[r]{$1$}}
\put(1416.0,877.0){\rule[-0.500pt]{4.818pt}{1.000pt}}
\put(220.0,113.0){\rule[-0.500pt]{1.000pt}{4.818pt}}
\put(220,68){\makebox(0,0){$10$}}
\put(220.0,857.0){\rule[-0.500pt]{1.000pt}{4.818pt}}
\put(342.0,113.0){\rule[-0.500pt]{1.000pt}{2.409pt}}
\put(342.0,867.0){\rule[-0.500pt]{1.000pt}{2.409pt}}
\put(413.0,113.0){\rule[-0.500pt]{1.000pt}{2.409pt}}
\put(413.0,867.0){\rule[-0.500pt]{1.000pt}{2.409pt}}
\put(464.0,113.0){\rule[-0.500pt]{1.000pt}{2.409pt}}
\put(464.0,867.0){\rule[-0.500pt]{1.000pt}{2.409pt}}
\put(503.0,113.0){\rule[-0.500pt]{1.000pt}{2.409pt}}
\put(503.0,867.0){\rule[-0.500pt]{1.000pt}{2.409pt}}
\put(535.0,113.0){\rule[-0.500pt]{1.000pt}{2.409pt}}
\put(535.0,867.0){\rule[-0.500pt]{1.000pt}{2.409pt}}
\put(563.0,113.0){\rule[-0.500pt]{1.000pt}{2.409pt}}
\put(563.0,867.0){\rule[-0.500pt]{1.000pt}{2.409pt}}
\put(586.0,113.0){\rule[-0.500pt]{1.000pt}{2.409pt}}
\put(586.0,867.0){\rule[-0.500pt]{1.000pt}{2.409pt}}
\put(607.0,113.0){\rule[-0.500pt]{1.000pt}{2.409pt}}
\put(607.0,867.0){\rule[-0.500pt]{1.000pt}{2.409pt}}
\put(625.0,113.0){\rule[-0.500pt]{1.000pt}{4.818pt}}
\put(625,68){\makebox(0,0){$100$}}
\put(625.0,857.0){\rule[-0.500pt]{1.000pt}{4.818pt}}
\put(747.0,113.0){\rule[-0.500pt]{1.000pt}{2.409pt}}
\put(747.0,867.0){\rule[-0.500pt]{1.000pt}{2.409pt}}
\put(819.0,113.0){\rule[-0.500pt]{1.000pt}{2.409pt}}
\put(819.0,867.0){\rule[-0.500pt]{1.000pt}{2.409pt}}
\put(869.0,113.0){\rule[-0.500pt]{1.000pt}{2.409pt}}
\put(869.0,867.0){\rule[-0.500pt]{1.000pt}{2.409pt}}
\put(909.0,113.0){\rule[-0.500pt]{1.000pt}{2.409pt}}
\put(909.0,867.0){\rule[-0.500pt]{1.000pt}{2.409pt}}
\put(941.0,113.0){\rule[-0.500pt]{1.000pt}{2.409pt}}
\put(941.0,867.0){\rule[-0.500pt]{1.000pt}{2.409pt}}
\put(968.0,113.0){\rule[-0.500pt]{1.000pt}{2.409pt}}
\put(968.0,867.0){\rule[-0.500pt]{1.000pt}{2.409pt}}
\put(991.0,113.0){\rule[-0.500pt]{1.000pt}{2.409pt}}
\put(991.0,867.0){\rule[-0.500pt]{1.000pt}{2.409pt}}
\put(1012.0,113.0){\rule[-0.500pt]{1.000pt}{2.409pt}}
\put(1012.0,867.0){\rule[-0.500pt]{1.000pt}{2.409pt}}
\put(1031.0,113.0){\rule[-0.500pt]{1.000pt}{4.818pt}}
\put(1031,68){\makebox(0,0){$1000$}}
\put(1031.0,857.0){\rule[-0.500pt]{1.000pt}{4.818pt}}
\put(1153.0,113.0){\rule[-0.500pt]{1.000pt}{2.409pt}}
\put(1153.0,867.0){\rule[-0.500pt]{1.000pt}{2.409pt}}
\put(1224.0,113.0){\rule[-0.500pt]{1.000pt}{2.409pt}}
\put(1224.0,867.0){\rule[-0.500pt]{1.000pt}{2.409pt}}
\put(1275.0,113.0){\rule[-0.500pt]{1.000pt}{2.409pt}}
\put(1275.0,867.0){\rule[-0.500pt]{1.000pt}{2.409pt}}
\put(1314.0,113.0){\rule[-0.500pt]{1.000pt}{2.409pt}}
\put(1314.0,867.0){\rule[-0.500pt]{1.000pt}{2.409pt}}
\put(1346.0,113.0){\rule[-0.500pt]{1.000pt}{2.409pt}}
\put(1346.0,867.0){\rule[-0.500pt]{1.000pt}{2.409pt}}
\put(1373.0,113.0){\rule[-0.500pt]{1.000pt}{2.409pt}}
\put(1373.0,867.0){\rule[-0.500pt]{1.000pt}{2.409pt}}
\put(1397.0,113.0){\rule[-0.500pt]{1.000pt}{2.409pt}}
\put(1397.0,867.0){\rule[-0.500pt]{1.000pt}{2.409pt}}
\put(1417.0,113.0){\rule[-0.500pt]{1.000pt}{2.409pt}}
\put(1417.0,867.0){\rule[-0.500pt]{1.000pt}{2.409pt}}
\put(1436.0,113.0){\rule[-0.500pt]{1.000pt}{4.818pt}}
\put(1436,68){\makebox(0,0){$10000$}}
\put(1436.0,857.0){\rule[-0.500pt]{1.000pt}{4.818pt}}
\put(220.0,113.0){\rule[-0.500pt]{292.934pt}{1.000pt}}
\put(1436.0,113.0){\rule[-0.500pt]{1.000pt}{184.048pt}}
\put(220.0,877.0){\rule[-0.500pt]{292.934pt}{1.000pt}}
\put(45,495){\makebox(0,0){$\langle l(t)\rangle$}}
\put(828,23){\makebox(0,0){$t$}}
\put(220.0,113.0){\rule[-0.500pt]{1.000pt}{184.048pt}}
\put(252,832){\raisebox{-.8pt}{\makebox(0,0){$\Diamond$}}}
\put(323,785){\raisebox{-.8pt}{\makebox(0,0){$\Diamond$}}}
\put(388,759){\raisebox{-.8pt}{\makebox(0,0){$\Diamond$}}}
\put(460,729){\raisebox{-.8pt}{\makebox(0,0){$\Diamond$}}}
\put(529,695){\raisebox{-.8pt}{\makebox(0,0){$\Diamond$}}}
\put(601,661){\raisebox{-.8pt}{\makebox(0,0){$\Diamond$}}}
\put(672,625){\raisebox{-.8pt}{\makebox(0,0){$\Diamond$}}}
\put(743,592){\raisebox{-.8pt}{\makebox(0,0){$\Diamond$}}}
\put(814,555){\raisebox{-.8pt}{\makebox(0,0){$\Diamond$}}}
\put(885,520){\raisebox{-.8pt}{\makebox(0,0){$\Diamond$}}}
\put(957,484){\raisebox{-.8pt}{\makebox(0,0){$\Diamond$}}}
\put(1028,449){\raisebox{-.8pt}{\makebox(0,0){$\Diamond$}}}
\put(1099,414){\raisebox{-.8pt}{\makebox(0,0){$\Diamond$}}}
\put(1171,381){\raisebox{-.8pt}{\makebox(0,0){$\Diamond$}}}
\put(1242,347){\raisebox{-.8pt}{\makebox(0,0){$\Diamond$}}}
\put(1314,313){\raisebox{-.8pt}{\makebox(0,0){$\Diamond$}}}
\put(1385,279){\raisebox{-.8pt}{\makebox(0,0){$\Diamond$}}}
\put(625,610){\usebox{\plotpoint}}
\multiput(625.00,607.67)(1.061,-0.500){756}{\rule{2.373pt}{0.120pt}}
\multiput(625.00,607.92)(806.075,-382.000){2}{\rule{1.187pt}{1.000pt}}
\end{picture}
}
\caption{The average length of a polygon side,
  for the random orientation fragmentation process.  Shown are $\langle
  l(t)\rangle$ \vs\ $t$ (diamonds) and a line of slope $-1/2$ (solid)
  for reference.}
\label{Fig-5}
\end{figure}
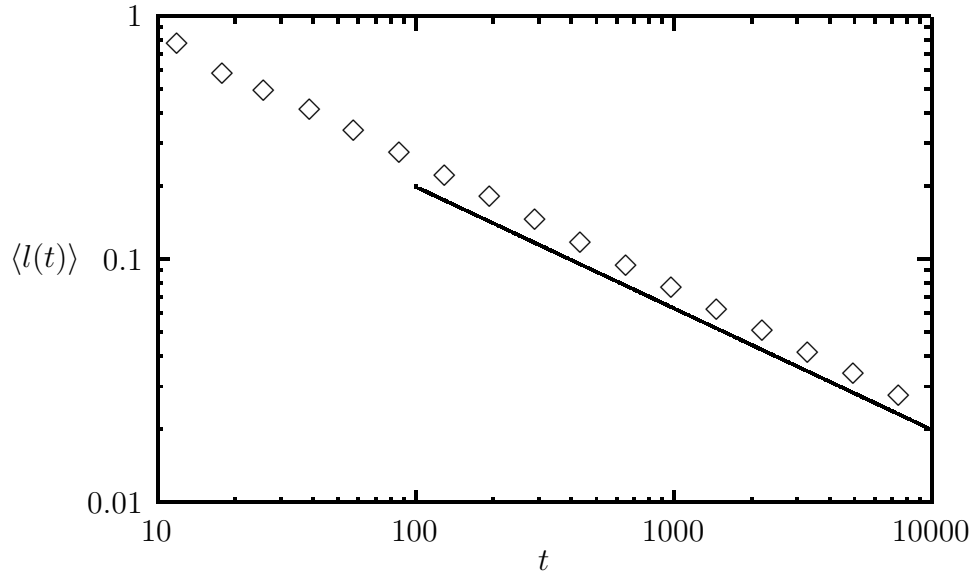

Intrigued by the rich kinetics of the rectangular fragmentation problem,
we also investigated numerically an isotropic fragmentation process. In
situations such as shattering of a thin glass plate or in membrane
crumpling, the fragments are polygons with a varying number of sides.
Hence, we introduce a process where a randomly oriented crack appears
with a uniform rate at a random point of the surface and propagates with
infinite speed until it meets existing cracks.  The original model can
be viewed as deposition of such perfectly oriented ``cross'' shaped cracks.
The overall fragmentation rate in both processes equals the volume of the
fragment. For randomly oriented fragmentation, each fragmentation
event creates an additional polygon and hence the total number
of polygons grows linearly in time according to $N(t)=M(1,1;t)=1+t$.
The average volume thus scales as $1/N(t)$ or $A\sim t^{-1}$.

A Monte-Carlo simulation study of isotropic fragmentation process
on a unit square suggests that unlike for oriented fragmentation,
only one length scale exists in the isotropic problem.
The average length of a polygon side is plotted in Fig.~4 and
appears to decay as $t^{-1/2}$. Therefore, the length follows the same
asymptotic behavior as does the square root of of the average area.
A snapshot of a realization of the system at time $t=1000$ is shown in
Fig.~5. This interesting picture suggests that it might prove insightful to
investigate various structure  properties of the system such as
the area distribution function and the side number probabilities
of the polygons.

In conclusion, we have studied two fragmentation processes in spatial
dimensions larger than one. For oriented fragmentation, where the
fragments are always rectangular, multiscaling is found in the long-time
limit. Specifically, the length distribution function has moments that
scale algebraically in time with an infinite number of independent
length scales, while the area distribution function is characterized by
a single length scale. The area distribution function also exhibits a
weak logarithmic singularity near the origin. Multiscaling appears to
depend strongly on the geometric nature of the process.
For isotropic fragmentation, a single length scale
describes the decay of the length as well as the area.


\bigskip\centerline{\bf Acknowledgment}\bigskip

After completing of this work we became aware of a closely related
study by D.~Boyer, G.~Tarjus and P.~Viot; we thank them for informing
us of their results.  We also want to thank S.~Cornell, E.~Kramer,
S.~Redner, and T.~Witten for interesting discussions.  We gratefully
acknowledge ARO grant \#DAAH04-93-G-0021, NSF grant \#DMR-9219845, and
the Donors of The Petroleum Research Fund, administered by the American
Chemical Society, for partial support of this research.




\bigskip\centerline{\bf Figure Captions}
\begin{itemize}
\begin{enumerate}
\item  Illustration of the fragmentation process in
                  two-dimensions.
\item  Realization of the fragmentation
                  process on a unit square at time $t=1000$.
\item The geometric solution. the hypersurface ${\bf s}^*$ satisfies
$\Pi_j s_j=2^d$.
\item
The average length of a polygon side,
                  for the random orientation fragmentation process.
                  Shown are $\langle l(t)\rangle$ \vs\ $t$ (diamonds) and a
                  line of slope $-1/2$ for reference.

\item   Realization of the random orientation fragmentation
                on a unit square at time $t=1000$.

\end{enumerate}
\end{itemize}
\end{document}